\documentclass[pra,showpacs,twocolumn,superscriptaddress]{revtex4-1}
\usepackage{amssymb}
\usepackage{amsmath}
\usepackage{graphicx}
\usepackage{epsfig}

\begin{document}

\title{Quench field sensitivity of two-particle correlation in a Hubbard
model}
\author{X. Z. Zhang}
\affiliation{School of Physics, Nankai University, Tianjin 300071, China}
\affiliation{College of Physics and Materials Science, Tianjin Normal University, Tianjin
300387, China}
\author{S. Lin}
\affiliation{School of Physics, Nankai University, Tianjin 300071, China}
\author{Z. Song}
\email{songtc@nankai.edu.cn}
\affiliation{School of Physics, Nankai University, Tianjin 300071, China}

\begin{abstract}
Short-range interaction can give rise to particle pairing with a short-range
correlation, which may be destroyed in the presence of an external field. We
study the transition between correlated and uncorrelated particle states in
the framework of one-dimensional Hubbard model driven by a field. We show
that the long time-scale transfer rate from an initial correlated state to
final uncorrelated particle states is sensitive to the quench field strength
and exhibits a periodic behavior. This process involves an irreversible
energy transfer from the field to particles, leading to a quantum
electrothermal effect.
\end{abstract}

\maketitle


\section{Introduction}

\label{sec_intro} The existence of short-range interaction can induce many
exotic physical phenomena, and therefore boost a theoretical interest in a
variety of correlated quantum systems. Owing to the rapid advance of
experimental techniques, ultracold quantum gases trapped in optical lattices
have provided in the past decade a route to simulate the physics of
different kinds of correlated quantum systems \cite{BlochNatP}. For few
interacting particles, experiments with ultracold atoms have so far
demonstrated the existence of bound pair (BP) states \cite{Winkler} and
correlated tunneling phenomena \cite{Folling}. This stimulate many
experimental and theoretical investigations aiming at the correlated
particles, which lead to a series of intriguing novel phenomena involving BP
formation \cite{Mahajan,Valiente1,MVExtendB,MVTB,Javanainen,Wang}, detection
\cite{Kuklov}, dynamics \cite%
{Petrosyan,Zollner,ChenS,Valiente2,Wang,JLNJP,JLBP,JLTrans}, and BP
condensate \cite{Rosch}.

On the other hand, recent experiments on optical lattices \cite{Greiner}
have renewed the interest for investigating the behavior of correlated
quantum system after a sudden change of Hamiltonian parameters, the
so-called quantum quench. The corresponding spectacular experimental results
have triggered an intensive research on quantum quenches in various systems
involving one-dimensional Bose systems \cite{Rigol}, Luttinger liquids \cite%
{Cazalilla} and others \cite{Manmana}. Among many remarkable quench
dynamical phenomena, it has been possible to observe collapse and revival of
matter waves with Bose-Einstein condensates in optical lattices \cite{Will},
coherent quench dynamics of fermionic atoms \cite{WillNC}, as well as
non-thermal behaviors in near-integrable experimental regimes \cite{Gring}.
A straightforward method to realize such quenched systems is the
instantaneous change of a global or local parameter of the system like an
external field or the interaction strength. In particular, when an external
field is applied to the correlated few-body system, preservation of the
correlation between the two particles in matter-wave transport can exhibit
unusual phenomena that are generally inaccessible in solid-state systems,
such as frequency doubling of Bloch oscillation (BO) of two correlated
particles \cite{Claro,Dias,Khomeriki,LonghiOL}, dynamic localization \cite%
{LonghiD}, coherent destruction of tunneling \cite{LonghiS} and the
fractional BO of atomic pairs \cite{Corrielli}. However, if the external
field is applied in a suitable manner, tunnelling between Bloch bands
becomes possible, which leads to a sudden death of the BO \cite{Lin} or
Bloch-Zener oscillation \cite{MUGA, Poladian, Greenberg, Kulishov,
Ruschhaupt, Longhi2012}.

In previous work \cite{Lin}, the dynamical behavior of BP has been
investigated for the extended Hubbard model driven by an external field. It
has been shown that the nearest-neighbor interaction can spoil the
completeness of the bound band, which can induce a sudden death of the
oscillation of BP and break the correlation between the particles when the
external field is switched on. However, the underlying mechanism of such
dynamical behaviors is not clear. Motivated by this question, here, we
theoretically investigate the underlying mechanism of the dynamics of the
two-particle correlation in the frame work of the Hubbard model driven by a
field, and show that, the presence of the external field can bring about an
avoided crossing between the correlated energy level and uncorrelated
scattering band leading to the correlation and energy transfer among
involved states. In this case, correlated particles can be pumped into the
scattering regime accompanied with the energy transfer from the field to
particles, rather than preserving the correlation of the particles in the
unmixed region in which the correlated and uncorrelated state are not mixed.
Intuitively, when an external field is applied, there can exist a threshold
below which the field cannot induce the destruction of the short-range
correlation between the particles. Furthermore, as the magnitude of the
field strength is beyond the threshold, the transfer rate from correlated
state to uncorrelated states gets larger with increase of the quench field
strength. However, in this paper, we show that the long time-scale transfer
rate from an initial correlated state to final uncorrelated particle states
exhibits a periodic behavior with the increase of the quench field strength,
which can occur in a small range of the field. The essential physics of such
counterintuitive dynamics is the periodic occurrence of the field-induced
multi avoided crossings between the correlated and uncorrelated energy
levels.

This paper is organized as follows. In Sec. \ref{sec_Machanism} we introduce
the model and discuss its dynamics in the avoided crossing situation through
a simple case. In Sec. \ref{sec_avoided}, an analysis of the two-particle
spectrum is presented, highlighting the existence of multi avoided
crossings. Sec. \ref{sec_dynamics} is devoted to investigate the quench
dynamics of two-particle correlation. Finally, in Sec. \ref{sec_Summary} the
main conclusions are outlined.

\begin{figure}[tbp]
\centering
\hskip-5.2mm 
\includegraphics[bb=37 17 596 468, width=0.5\textwidth, clip]{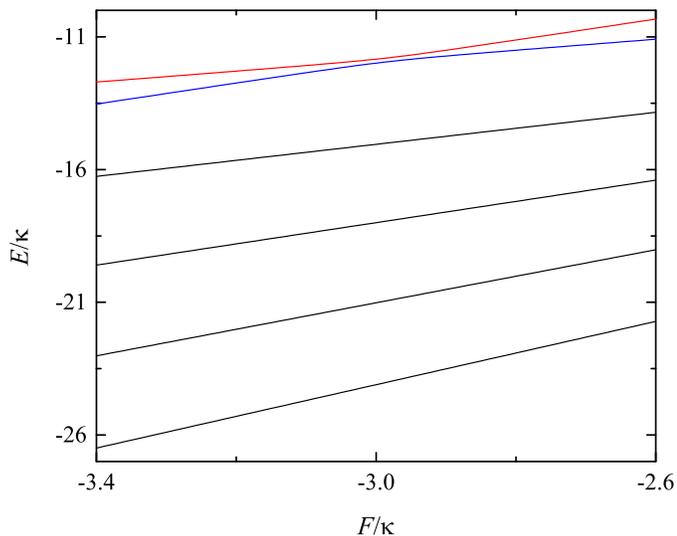}
\caption{(Color online) Two-particle energy spectrum of the $N=3$ system as a
function of parameter $F$. The other parameter values of the system are $U=V=-6,$ and $\kappa =0.4$. The avoided crossing involves the top two energy
levels denoted by the red and blue line, respectively, which occurs in the
vicinity of $F=U/2=-3$.} \label{fig1}
\end{figure}


\section{Model Hamiltonian and the mechanism}

\label{sec_Machanism} For the sake of generality and simplicity, we start
from a driven one dimensional extended Hubbard model. It is worth pointing
out that the results shown in the following are also applicable to the other
form of short-range interaction. The model concerned can be employed to
describe the dynamics of interacting particles on a one-dimensional
tight-binding lattice driven by an external field, which can be defined by
the Hamiltonian
\begin{equation}
H=H_{0}+F\sum_{j=1}jn_{j},
\end{equation}%
where the second term describes the linear external field served as a quench
parameter. And $H_{0}$\ is the one-dimensional Hamiltonian for the extended
Hubbard model on an $N$-site lattice%
\begin{eqnarray}
H_{0} &=&-\kappa \sum_{j=1}\left( a_{j}^{\dagger }a_{j+1}+\text{\text{H.c.}}%
\right) +\frac{U}{2}\sum_{j=1}n_{j}\left( n_{j}-1\right)  \notag \\
&&+V\sum_{j=1}n_{j}n_{j+1},  \label{H0}
\end{eqnarray}%
where $a_{j}^{\dag }$ ($a_{j}$) is the creation (annihilation) operator for
boson at site $j$ and $n_{j}=a_{j}^{\dag }a_{j}$ is the number operator. $F$
is magnitude of an external field. The tunneling amplitude and on-site
interaction strength are denoted by $\kappa $ and $U$, respectively. And $V$
accounts for the nearest-neighbor (NN) interaction. The Hamiltonian (\ref{H0}%
) can also describe ultracold atoms or molecules with magnetic or electric
dipole-dipole interactions in optical lattices \cite{Zoller2008}. In the
absence of the external field, the short-range interaction denoted by $U$\
and $V$ can induce the particle pairing with short-range correlation no
matter what magnitude of the particle-particle interaction is \cite{Lin}. We
consider a correlated particles state. The spectrum of the system after a
field quench is dramatically changed, which may destroy the correlation
between the particles. We will demonstrate this point through a simple case.


\begin{figure}[tbp]
\centering\hskip-5.2mm 
\includegraphics[bb=44 21 596 468, width=0.5\textwidth, clip]{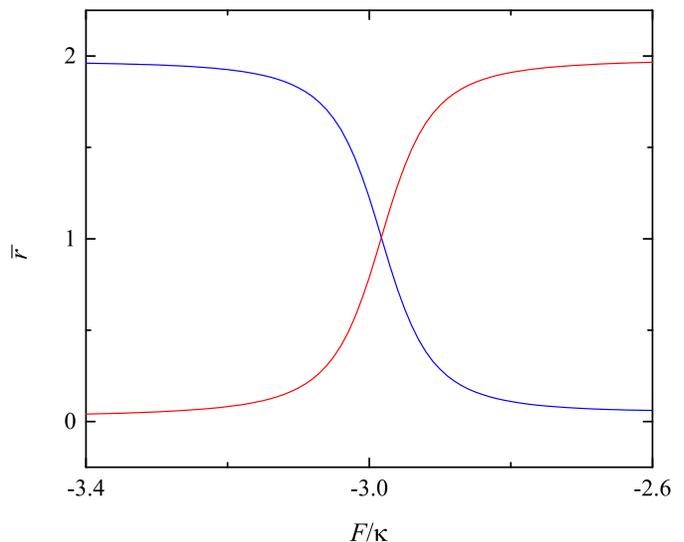}
\caption{(Color online) Plot of the relative distance $\overline{r}\
$ as a function of parameter $F$. The red and blue lines denote the top two
energy levels of the system shown in Fig. \ref{fig1}. It can be observed that the
two corresponding states exchange the correlation through an avoided
crossing region.} \label{fig2}
\end{figure}

We first focus on 3-site chain, which possess a simple energy structure but
can reveal the underlying mechanism of the field-induced destruction of the
correlation between the particles. The energy levels of the concerned system
as a function of $F$ with the parameter $U=V=-6$ is plotted in Fig. \ref%
{fig1}. It is shown that there are six energy levels. And we focus on the
top two energy levels, which undergo an avoided crossing at the point of $%
U=V=2F$. When the parameters satisfy the condition $\left\vert
F-U/2\right\vert \ll U=V$ and $F$, $U\gg \kappa $, the dynamics of the
correlated particles involving top two energy levels is governed by the
following effective Hamiltonian
\begin{eqnarray}
H_{\text{eff}} &=&\frac{1}{2}\left( \frac{\sqrt{2}\kappa ^{2}}{U-F-V}+\frac{%
\sqrt{2}\kappa ^{2}}{F-V}\right) \left\vert \text{up}\right\rangle
\left\langle \text{p}\right\vert +\text{H.c.}  \notag \\
&&+\left( 4F+\frac{2\kappa ^{2}V}{F^{2}-V^{2}}\right) \left\vert \text{up}%
\right\rangle \left\langle \text{up}\right\vert  \notag \\
&&+\left( U+2F+\frac{2\kappa ^{2}}{U-V-F}\right) \left\vert \text{p}%
\right\rangle \left\langle \text{p}\right\vert \text{,}
\end{eqnarray}%
which is obtained by the canonical transformation \cite{Nakajima}. And $%
\left\vert \text{p}\right\rangle =a_{1}^{\dagger 2}/\sqrt{2}\left\vert \text{%
vac}\right\rangle $ and $\left\vert \text{up}\right\rangle =a_{1}^{\dagger
}a_{3}^{\dagger }\left\vert \text{vac}\right\rangle $ denote pair and unpair
state, respectively. In order to investigate the two-particle correlation of
given state, we introduce the average distance between the two particles
defined as
\begin{equation}
\overline{r}=\sum_{i,r}r\left\langle \Psi \right\vert n_{i}n_{i+r}\left\vert
\Psi \right\rangle ,
\end{equation}%
which can measure the correlation between the two particles. We plot the
average distance $\overline{r}$ as a function of $F$ in Fig. \ref{fig2}, in
which the blue and red line denote the correlation of top two energy levels,
respectively. It can be seen from that when the external field $F$ varies to
the vicinity of the point $U=V=2F$, the two states are mixed associated with
the exchange of correlation through the avoided crossing. Based on this, one
can assume that the occurrence of the avoided region can lead the correlated
particles to displaying a distinct dynamical behavior for the system with
and without avoided crossing. To demonstrate this point, we consider the
time evolution of the initial state $\left\vert \Psi \left( 0\right)
\right\rangle =\left\vert \text{p}\right\rangle $, which can be readily
obtained as%
\begin{eqnarray}
\left\vert \Psi \left( t\right) \right\rangle &=&e^{-iC_{0}t}\left\{ \left[
\cos \left( \omega t\right) +i\cos \theta \sin \left( \omega t\right) \right]
\left\vert \text{p}\right\rangle \right.  \notag \\
&&\left. -\left[ i\sin \theta \sin \left( \omega t\right) \right] \left\vert
\text{up}\right\rangle \right\} ,
\end{eqnarray}%
where
\begin{eqnarray}
C_{0} &=&\frac{U}{2}+3F+\frac{\kappa ^{2}U}{F^{2}-U^{2}}-\frac{\kappa ^{2}}{F%
}, \\
C_{1} &=&U-2F-\frac{2\kappa ^{2}}{F}-\frac{2\kappa ^{2}U}{F^{2}-U^{2}}, \\
\omega &=&\frac{1}{2}\sqrt{\left( \frac{U\sqrt{2}\kappa ^{2}}{F^{2}-FU}%
\right) ^{2}+C_{1}^{2}}, \\
\tan \theta &=&\frac{U\sqrt{2}\kappa ^{2}}{C_{1}\left( F^{2}-FU\right) }.
\end{eqnarray}%
Here we focus on how much probability of the evolved state $\left\vert \Psi
\left( t\right) \right\rangle $ will remain in the state $\left\vert \text{p}%
\right\rangle $ as time goes on. To this aim, we define the\ transfer rate
from the initial paired state to the final unpaired state as
\begin{equation}
P\left( t\right) =1-\left\vert \left\langle \text{p}\right. \left\vert \Psi
\left( t\right) \right\rangle \right\vert ^{2}=\sin ^{2}\theta \sin
^{2}\left( \omega t\right) .
\end{equation}%
%
%
%
%
%
%
%
%
%
%
%
%
%
%
%
%
%
%
%
%
%
%
%
%
%
%
%
%
%
%
%
%
%
%
%
%
%
%
%
%
%
%
%
%
%
%
%
%

\begin{figure}[tbp]
\centering\hskip-5.2mm 
\includegraphics[bb=23 21 596 468, width=0.5\textwidth, clip]{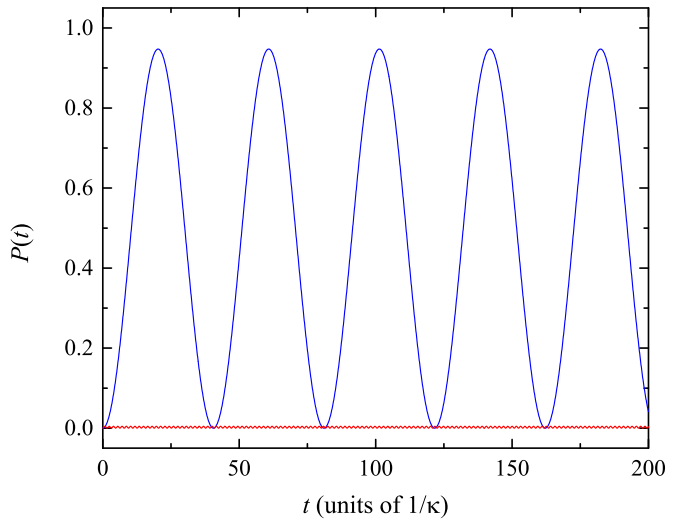}
\caption{(Color online) Evolution of the transfer probability $P\left( t\right) $.
The other system parameters are the same as that in Fig. \ref{fig1}. The initial
state is prepared in the state $\left\vert \text{p}\right\rangle $. The blue
and red lines represent evolutions of $P\left( t\right) $ with the two
typical values of $F=-3$, $-1$, which is corresponding to the system with
and without the avoided crossing, respectively.} \label{fig3}
\end{figure}

For the case of $F=U/2$ that the parameter is in the avoided crossing
region, we have $C_{1}=-4\kappa ^{2}/3U$, $C_{0}=4F-8\kappa ^{2}/3U$, $%
\omega =\sqrt{76}\kappa ^{2}/3\left\vert U\right\vert $, and $\tan \theta =3%
\sqrt{2}$, which lead to $P\left( t\right) =18\sin ^{2}\left( \sqrt{76}%
\kappa ^{2}t/3\left\vert U\right\vert \right) /19$. The transfer rate as a
function of time is plotted in Fig. \ref{fig3}. It can be shown that the
transition between paired and unpaired state occurs with period $%
t_{r}=3\left\vert U\right\vert \pi /\sqrt{76}\kappa ^{2}$, which is inverse
proportion to the energy gap between the involved two energy levels. In
their essence, the mixing of the pair and unpair state leads to the exchange
of their correlation in the region of avoided crossing. On the other hand,
when the parameter is away from the avoided crossing region satisfying the
condition of $\left\vert F-U/2\right\vert \gg \kappa $, we have $%
C_{1}=U-2F=2\omega $, $C_{0}=U/2+3F$, $\sin \theta =\tan \theta =U\sqrt{2}%
\kappa ^{2}/C_{1}\left( F^{2}-FU\right) $, which yield $P\left( t\right) =%
\left[ U\sqrt{2}\kappa ^{2}/C_{1}\left( F^{2}-FU\right) \right] ^{2}\sin ^{2}%
\left[ \left( U-2F\right) t/2\right] \approx 0$ owing to the condition of $%
\left\vert F-U/2\right\vert \gg \kappa $. This indicates that the evolved
state will remain in the paired state rather than tunneling to the unpaired
state as shown in Fig. \ref{fig3}. Through the above analysis, it is clear
that the existence of the avoided crossing is crucial for the coherent
motion of the correlated particles. However, the structure of the avoided
crossing for large $N$ system will be something different comparing with
that of the above $N=3$ case. In the following section, we will investigate
the large $N$ system.

\begin{figure}[tbp]

\includegraphics[bb=0 15 596 500,width=0.42\textwidth,
clip]{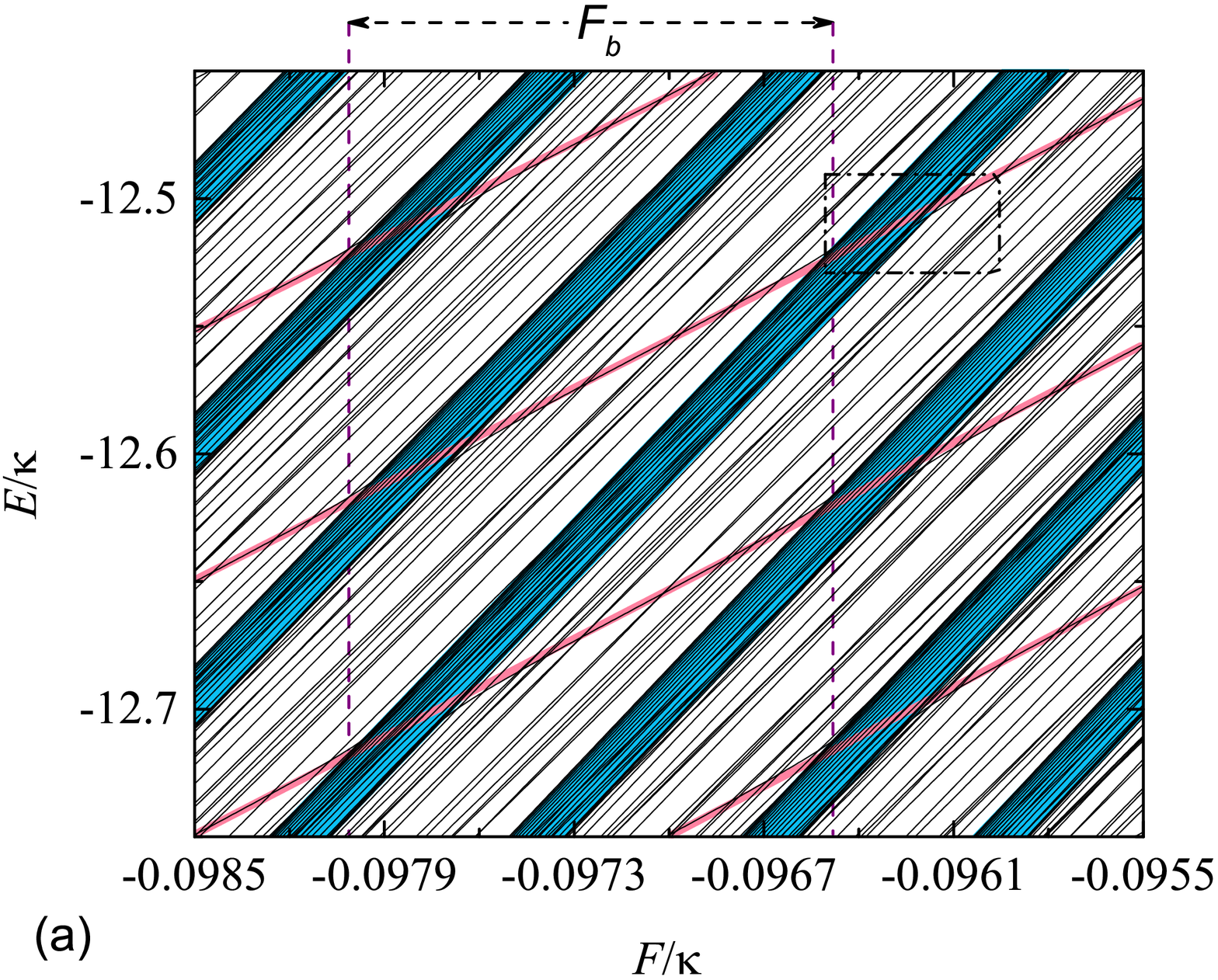}
\includegraphics[bb=0 15 596 500,width=0.42\textwidth,
clip]{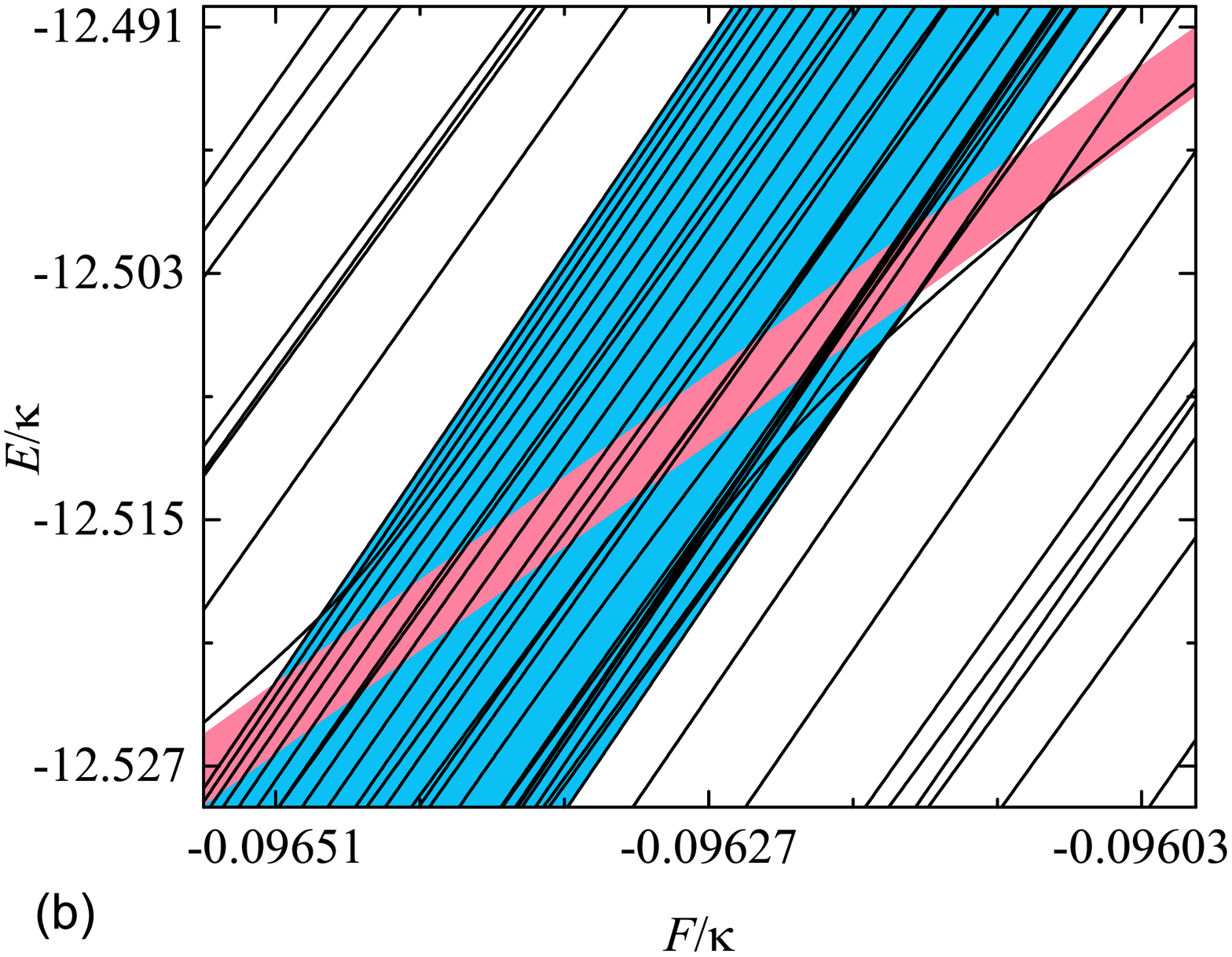}
\includegraphics[bb=0 15 596 500,width=0.42\textwidth,
clip]{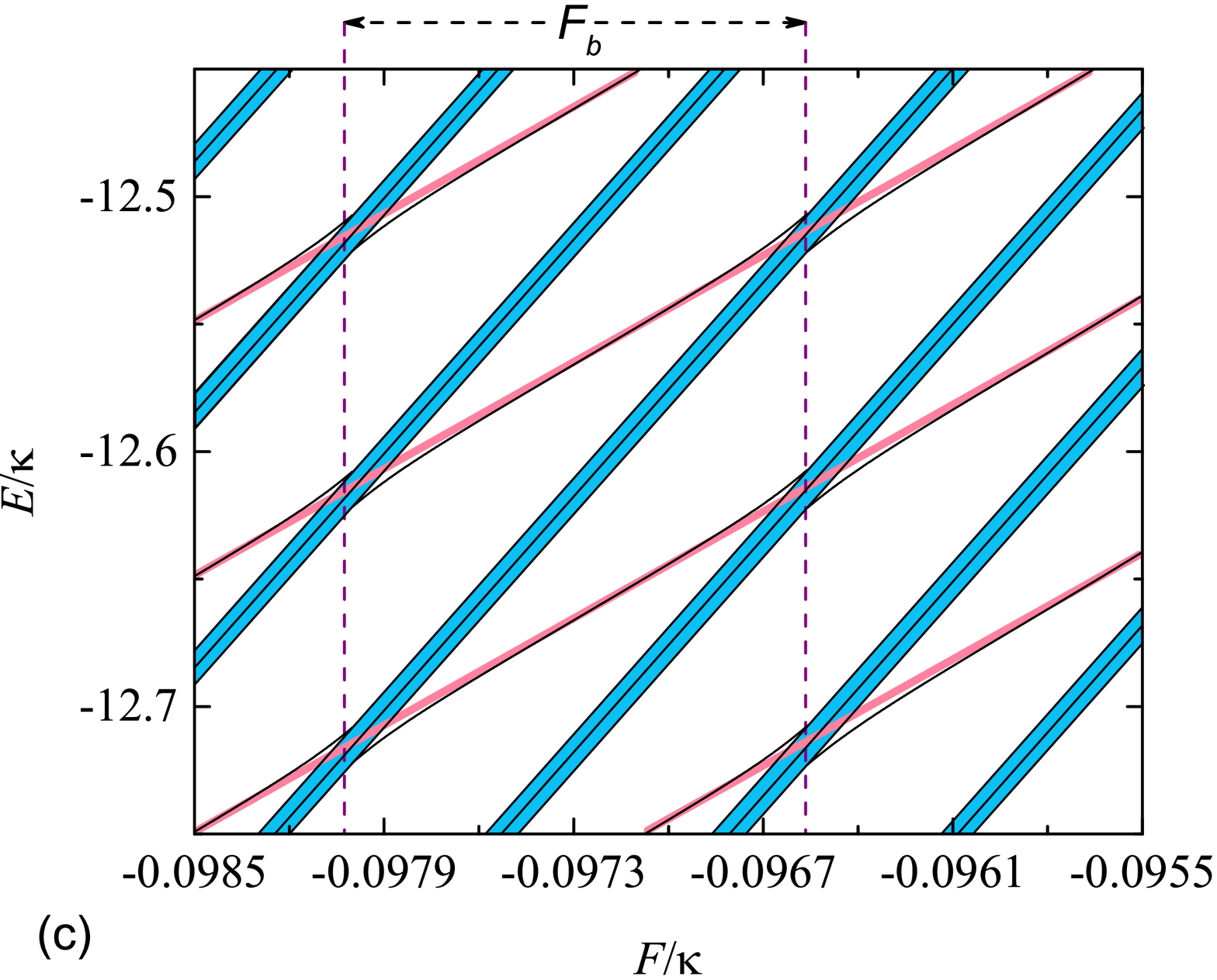}
\caption{(Color online) Energy spectrum of the two correlated particles as a
function of the external field $F$ for the system with $U=V=-6.24$, $\protect\kappa =1$. The blue and red region indicates correlated energy level and
multi uncorrelated energy levels, respectively. It can be observed that a
slight variation of the external field can give rise to the periodic
occurrence of the avoided crossing region, in which the correlated state and
multi uncorrelated states are mixed together. The black dashed rectangle denotes a typical avoided crossing. (b) is detail with enlarged
scale of black dashed marker in (a). (c) Schematic illustration of the ideal
avoided crossing regions. Each of these contains three scattering and two
correlated energy levels. It should be noted that the slopes of the two adjacent
correlated lines are slightly different, which is estimated about $F$.
This conclusion can also be applied to the adjacent uncorrelated bands
composed by multi scattering energy levels.} \label{fig4}
\end{figure}


\section{Energy spectrum and multi avoided crossings}

\label{sec_avoided} To investigate the influence of an external field on the
energy band structure, we first give a two-particle solution of the
Hamiltonian $H_{0}$. For the sake of simplicity, we restrict the analysis to
the system with $U=V<0$, which ensures the Bloch band for BP locates below
the continuum of scattering states. As in Refs. \cite{JLBP,Lin}, a state in
the two-particle Hilbert space, can be expanded in the basis set $\left\{
\left\vert \phi _{r}^{K}\right\rangle ,r=0,1,2,...\right\} $, with%
\begin{eqnarray}
\left\vert \phi _{0}^{K}\right\rangle &=&\frac{1}{\sqrt{2N}}%
\sum_{j}e^{iKj}\left( a_{j}^{\dag }\right) ^{2}\left\vert \text{\text{\text{%
vac}}}\right\rangle , \\
\left\vert \phi _{r}^{K}\right\rangle &=&\frac{1}{\sqrt{N}}%
e^{iKr/2}\sum_{j}e^{iKj}a_{j}^{\dag }a_{j+r}^{\dag }\left\vert \text{\text{%
vac}}\right\rangle ,
\end{eqnarray}%
where $\left\vert \text{vac}\right\rangle $\ is the vacuum state for the
boson operator $a_{i}$. Here $K$ denotes the center momentum of two
particles, and $r$ is the distance between the two particles. Due to the
translational symmetry of the present system, we have the following
equivalent Hamiltonian:

\begin{eqnarray}
H_{\text{eq}}^{K} &=&-J_{K}(\sqrt{2}\left\vert \phi _{0}^{K}\right\rangle
\left\langle \phi _{1}^{K}\right\vert +\sum_{j=1}\left\vert \phi
_{j}^{K}\right\rangle \left\langle \phi _{j+1}^{K}\right\vert +\text{\text{%
H.c.}})  \notag \\
&&+U\left( \left\vert \phi _{0}^{K}\right\rangle \left\langle \phi
_{0}^{K}\right\vert +\left\vert \phi _{1}^{K}\right\rangle \left\langle \phi
_{1}^{K}\right\vert \right)  \label{H_eq}
\end{eqnarray}%
in each invariant subspace indexed by $K$. In its present form, $H_{\text{eq}%
}^{K}$ are formally analogous to the tight-binding model describing a
single-particle dynamics in a semi-infinite chain with the $K$-dependnet
hopping integral $J_{K}=2\kappa \cos \left( K/2\right) $ in thermodynamic
limit $N\rightarrow \infty $. In this paper, we are interested in the BP
states, which corresponds to the bound state solution of the single-particle
Schr\"{o}dinger equation

\begin{equation}
H_{\text{eq}}^{K}\left\vert \psi _{K}\right\rangle =\epsilon _{K}\left\vert
\psi _{K}\right\rangle .
\end{equation}%
For a given $J_{K}$, the Hamiltonian $H_{\text{eq}}^{K}$ possesses one or
two types of bound states \cite{Lin}, which are denoted as $\left\vert \psi
_{K}^{+}\right\rangle $\ and $\left\vert \psi _{K}^{-}\right\rangle $,
respectively. Here the Bethe-ansatz wave functions have the form%
\begin{equation}
\left\vert \psi _{K}^{\pm }\right\rangle =C_{0}^{K}\left\vert \phi
_{0}^{K}\right\rangle +\left( \pm 1\right) ^{r}C_{r}^{K}e^{-\beta
r}\left\vert \phi _{r}^{K}\right\rangle ,
\end{equation}%
with $\beta >0$. For two such bound states $\left\vert \psi _{K}^{\pm
}\right\rangle $ the Schr\"{o}dinger equation in Eq. (\ref{H_eq}) admits%
\begin{equation}
\pm e^{3\beta }+2u_{K}e^{2\beta }\pm \left( u_{K}^{2}-1\right) e^{\beta
}+u_{K}=0,
\end{equation}%
where $u_{K}=U/J_{K}$ is the reduced interaction strength. The corresponding
bound-state energy of $\left\vert \psi _{K}^{\pm }\right\rangle $ can be
expressed as\
\begin{equation}
\epsilon _{K}^{\pm }=\pm J_{K}\cosh \beta .
\end{equation}%
As is shown in Ref. \cite{Lin}, if $\left\vert U/\kappa \right\vert $ is
larger than $6$, the system comprises two complete bands, one of which
corresponds to two particles bound states. Otherwise, it possesses the
incomplete bound band associated with the vanishing of the gap between the
scattering band and bound band. Here we want to stress that the preservation
of the correlation between the particles in transport does not depend on the
completeness of bound band of $H_{0}$\ as the external field is switched on.
The dynamics of the two correlated particles is determined by the existence
of the avoided crossings of $H$, which can destroy the pair correlation. We
will demonstrate this point in the following section.

Now, we switch gears to the spectrum of the Hamiltonian $H$. For the case of
$U\gg F$, $\kappa $, the energy spectrum of the two correlated particles
becomes discrete and turns out to be composed by two separated Wannier-Stark
ladders \cite{Longhi2012}. In this case, the BP undergoes a perfect BO with
period $t_{B}=2\pi /F$ and\ preserves the pair correlation, which is
analogous to the case of a single particle motion subjected to an external
field \cite{LonghiOL,Longhi2012}. On the other hand, the presence of the
external field can not only widen the spectrum of the scattering and bound
band of $H_{0}$ but also submerge the bound band into the scattering band.
In this case, surprisingly, we find that the correlated energy level can
overlap with the uncorrelated band formed by multi uncorrelated energy
levels. This can be seen from Fig. \ref{fig4}(a), in which the blue
correlated energy level overlaps with the first red uncorrelated band, then
in turn encounters the different uncorrelated band forming multi avoided
regions. In these regions, the corresponding wave functions of two types of
energy levels are mixed accompanied with the correlation and energy exchange
among the involved states, which can prevent the correlated transport of the
two particles. It is worth pointing out that the existence of multi avoided
crossing regions is sensitive to the external field $F$\ and presents a
periodic behavior with period $F_{b}=0.0015$\ in such a small range of the
external field as shown in Fig. \ref{fig4}(a) and (c). For the sake of
clarity and simplicity, we draw schematically the ideal avoided crossing
regions between correlated energy level and scattering band and denote the
period of the occurrence of these regions in\quad Fig. \ref{fig4}(c). One
should be noted that the slope of the two adjacent blue correlated lines are
different and so are the red uncorrelated bands, which cannot affect the
periodic occurrence of the avoided crossing regions in such a small range of
the external field $F$. It is presumable that the dramatic change of the
energy spectrum from avoided region to unmixed region induced by a slight
variation of $F$ can be characterized by the dynamics of correlated
particles and exhibits some peculiar behaviors, which will be investigated
in the following section.

\begin{figure}[tbp]
\centering\hskip-5.2mm 
\includegraphics[bb=23 171 593 602, width=0.5\textwidth, clip]{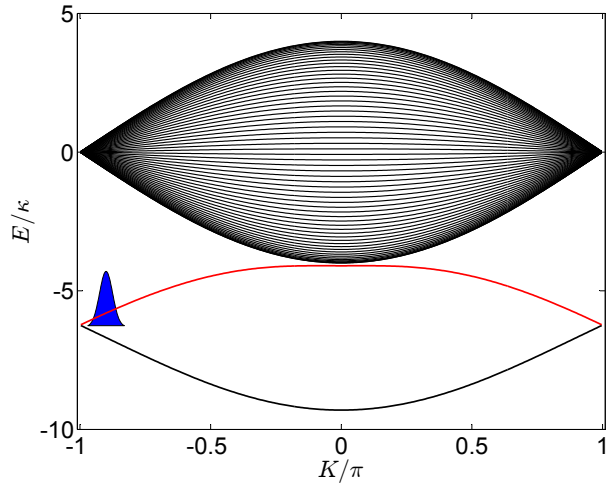}
\caption{(Color online) Schematics of the initial correlated wave packet
projected in the energy space of $H_{0}$. The blue Gaussian wave packet is
centered around $K_{0}=-0.9\pi $ of the upper bound band denoted by the red
line.} \label{fig5}
\end{figure}


\begin{figure}[htbp]
\centering\hskip-3.2mm
\includegraphics[bb=0 20 596 465,width=0.5\textwidth,
clip]{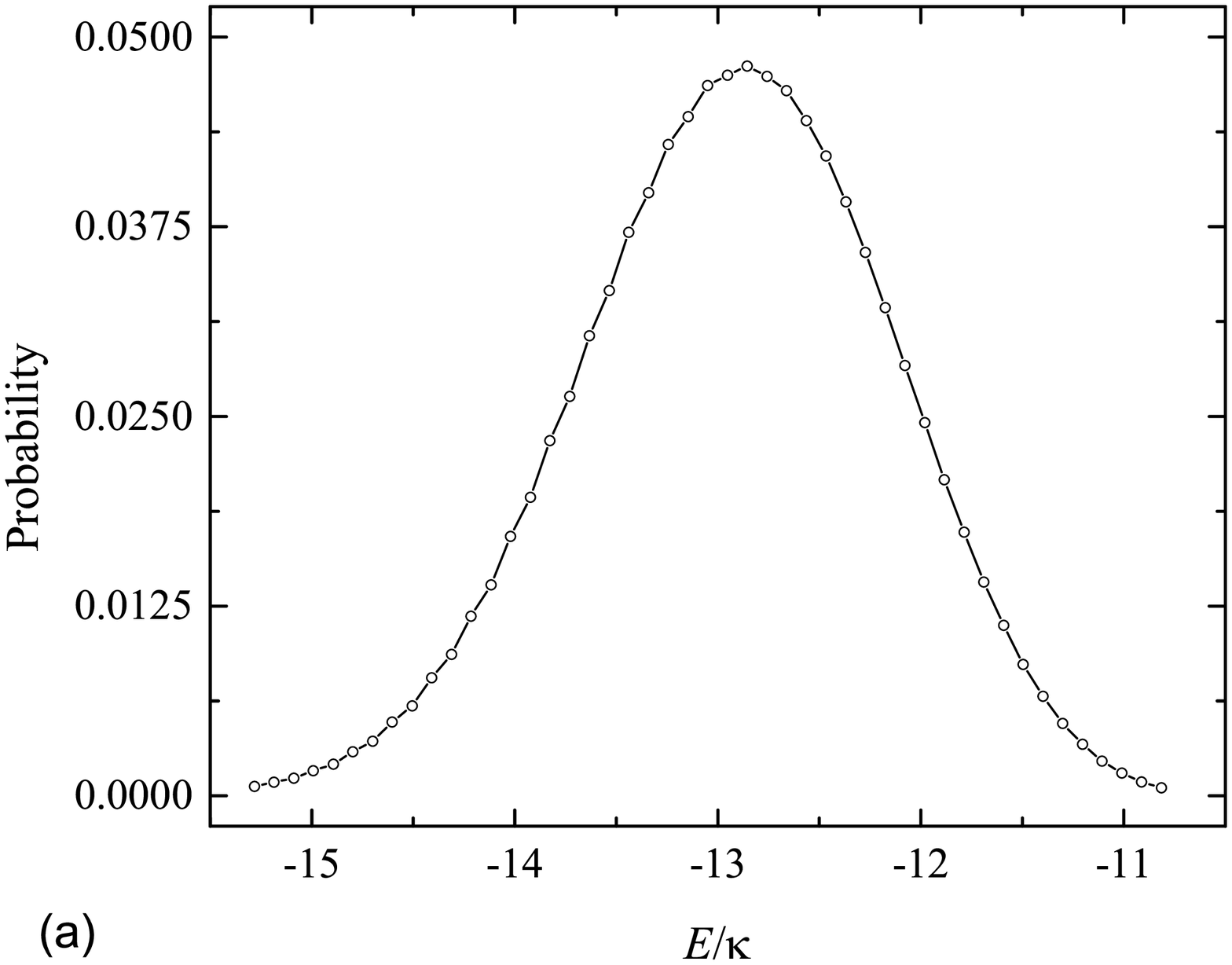} 
\includegraphics[bb=6 20 596 500,width=0.5\textwidth,
clip]{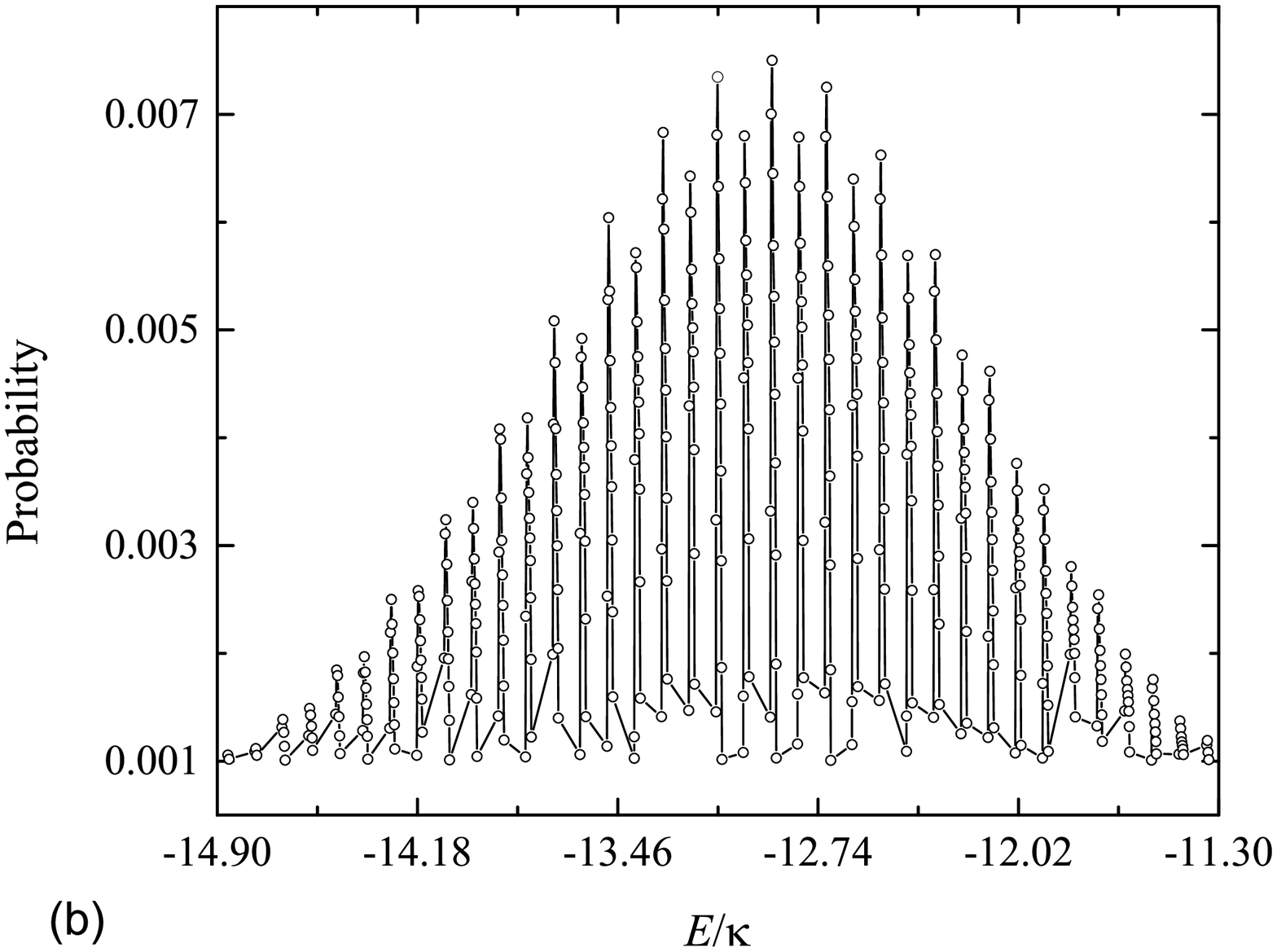}
\caption{(Color online) The main distribution of the initial BP wave packet in the
energy space for the systems with parameters (a) $F=-0.097120,$ and (b) $F=-0.097815$, respectively. The parameters $U=V=-6.24,$ $\kappa =1$ and $N=111$ are the same for both systems. It is should be noted that we restrict
the total probability of the initial state distributed on the energy space
to be $0.95$, which can characterize the main distribution of the initial
state. In (a) the profile of energy distribution of the initial state
exhibits Gaussian-like form and involves less energy. (b) denotes the distribution in the system with multi avoided crossing regions involving more energy levels comparing with (a). The interval between the two adjacent energies related to either correlated states or multi scattering states is about $F$, which agrees with our prediction.}
\label{fig6}
\end{figure}


\section{Quench dynamics of two-particle correlation}

\label{sec_dynamics}
\begin{figure*}[tbp]
\centering
\hskip -3.2mm
\includegraphics[bb=0 43 542 749,width=0.47\textwidth,
clip]{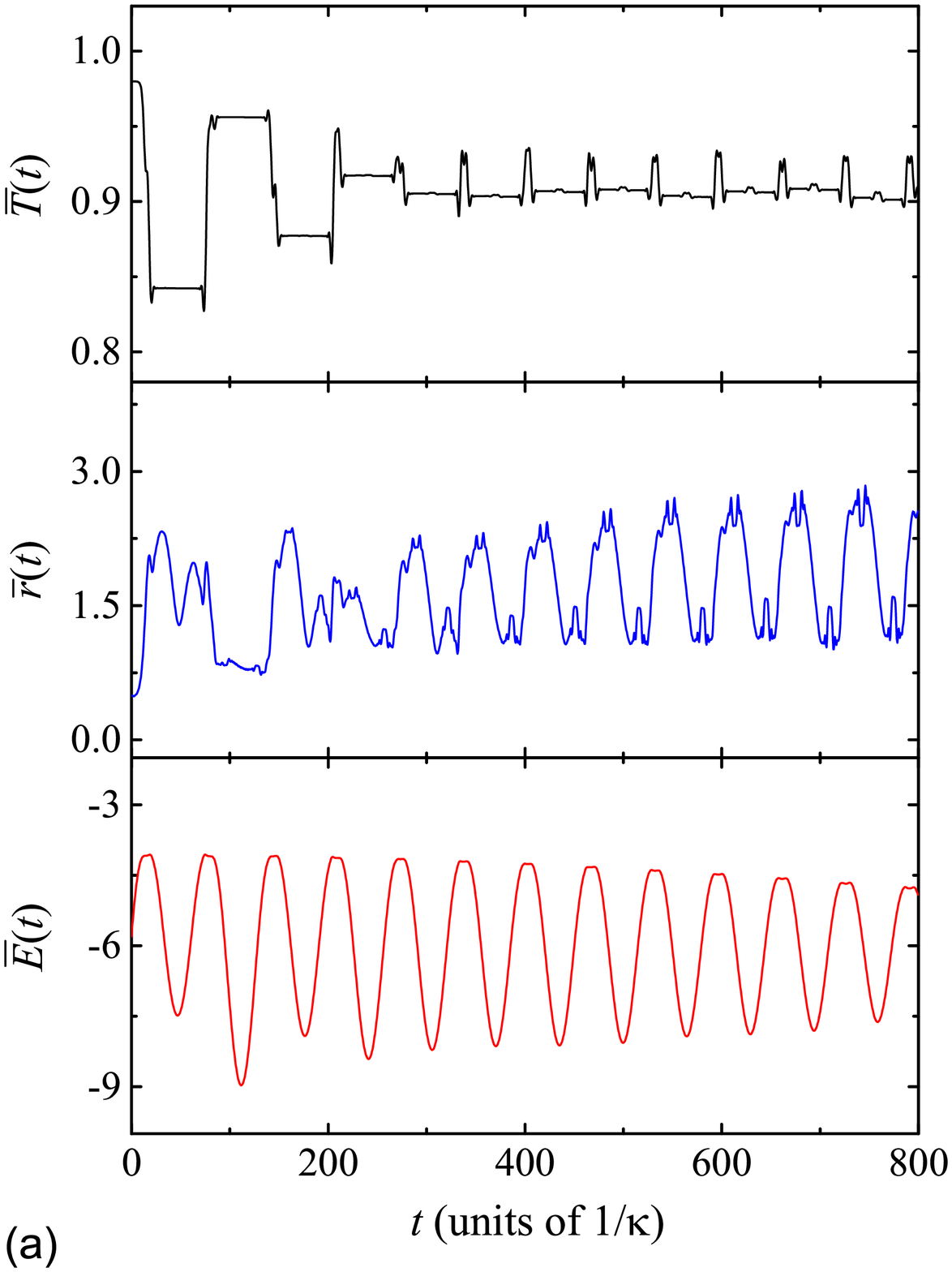} 
\includegraphics[bb=0 43 542 749,width=0.47\textwidth,
clip]{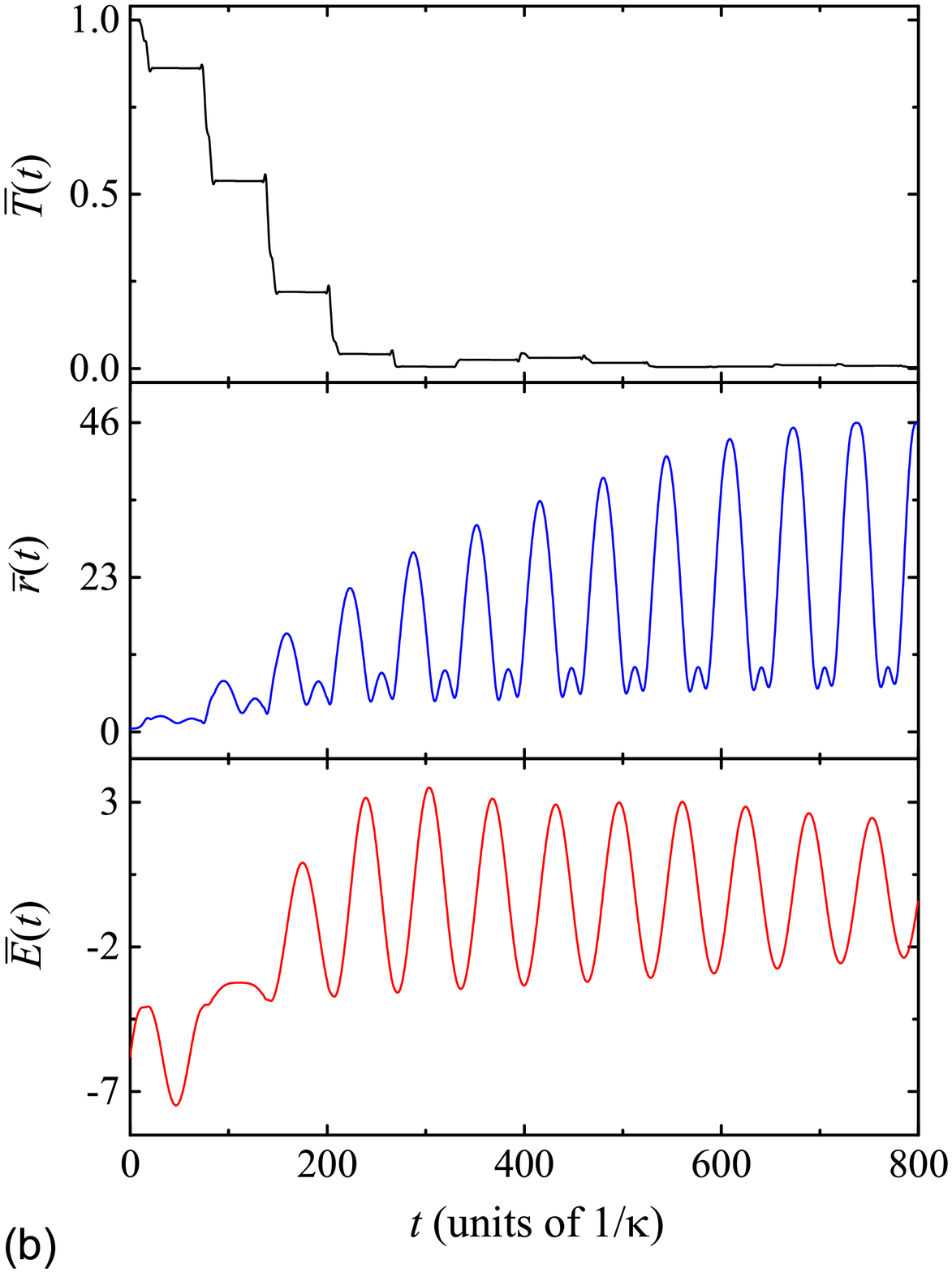}
\caption{(Color online) Numerically computed evolution of the functions $%
\overline{T}\left( t\right) $, $\overline{r}\left( t\right) $, and $%
\overline{E}\left( t\right) $ for the initial BP wave packet in the form of
Eq. (\protect\ref{BP_wp}) with parameter values given in the text and for
two typical values of $F$: (a) $F=-0.097120,$ and (b) $F=-0.097815$. The
other parameter values are $U=V=-6.24,$ $\protect\kappa =1$ and $N=111$. (a)
shows that the evolved state remains its correlation and preserves the
constancy of the average interval energy. This denotes that the two
correlated particles undergoes a BO with period $t_{B}\approx 65$. While in
(b), the increase of $\overline{r}\left( t\right) $, $\overline{E}\left(
t\right) $ and decrease of $\overline{T}\left( t\right) $ are associated
with the Zener tunneling from the bound band to the scattering band, which
leads to the dissociation of the two-particle correlation. }
\label{fig7}
\end{figure*}

\begin{figure}[tbp]
\centering\hskip-7.2mm 
\includegraphics[bb=27 23 576 500, width=0.5\textwidth, clip]{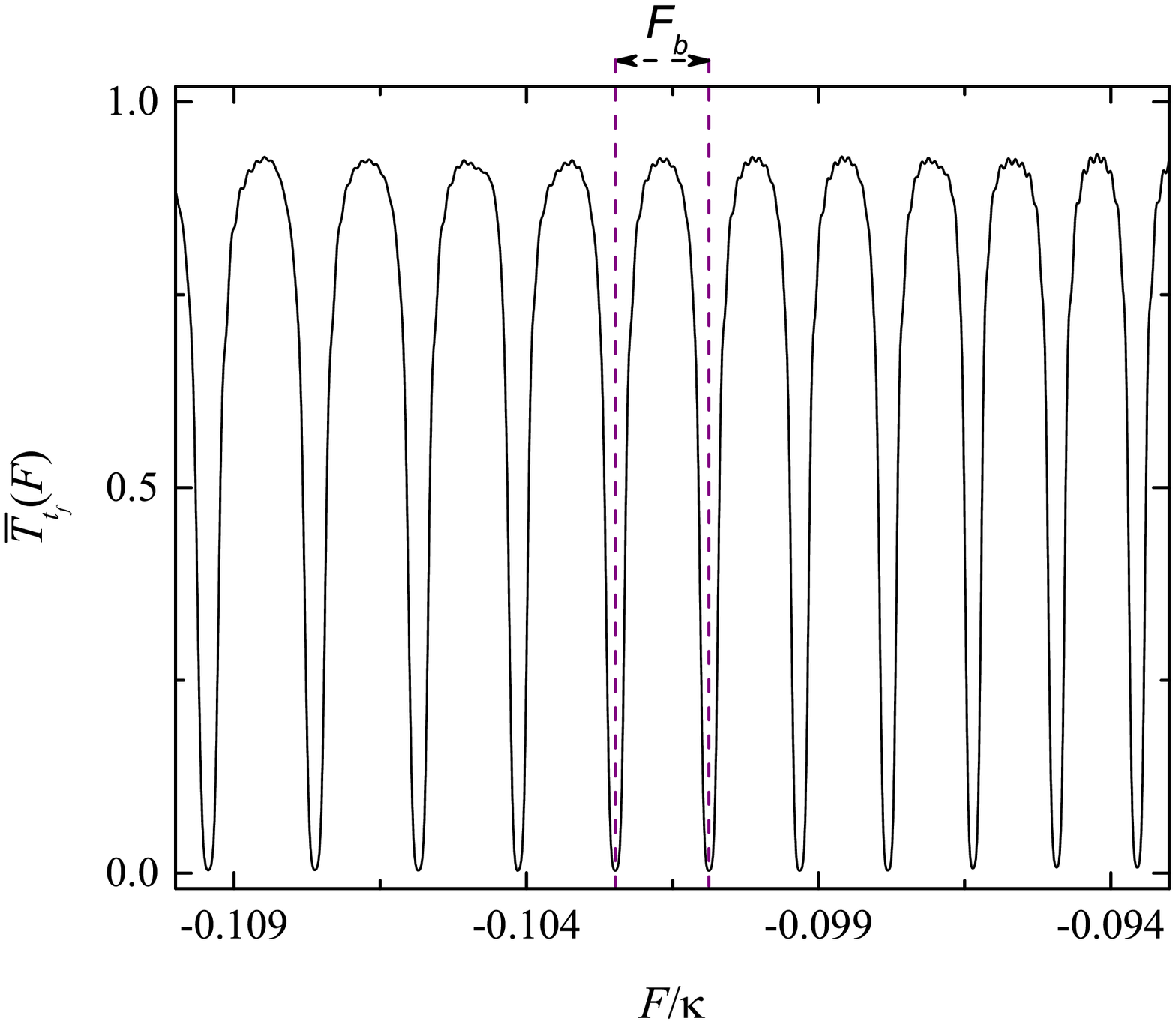}
\caption{(Color online) Behavior of the function $\overline{T}_{t_{f}}\left(
F\right) $ versus the field strength $F$ for the parameter $t_{f}=800$. The
other parameter values are the same with that of Fig. \ref{fig4}. Note that the $\overline{T}_{t_{f}}\left( F\right) $ is periodic with period $F_{B}=0.0015$, which accords with the period of periodic occurrence of the multi avoided
crossings in the system.} \label{fig8}
\end{figure}

To gain insights into the transition of the spectrum induced by the external
field. We perform a numerical simulation to investigate the dynamical
behaviors of correlated particles. The initial correlated particles state we
concerned is placed on the upper bound band of $H_{0}$,
\begin{equation}
\left\vert \Psi \left( 0\right) \right\rangle =\Lambda \sum_{K}\exp \left[ -%
\frac{\left( K-K_{0}\right) ^{2}}{2\alpha ^{2}}-iN_{A}K\right] \left\vert
\psi _{K}\right\rangle ,  \label{BP_wp}
\end{equation}%
where $\Lambda $ is the normalization factor, $K_{0}$ and $N_{A}$ denote the
central momentum and position of the initial wave packet, respectively.
Throughout this paper, we set the parameter $K_{0}=-0.9\pi $, $\alpha =0.2$,
and $N_{A}=36$. For the sake of clarity, we schematically illustrate the
initial state $\left\vert \Psi \left( 0\right) \right\rangle $ in Fig. \ref%
{fig5}. We consider a quantum quench in which the external field strength is
suddenly changed from $F=0$\ for $t=0$\ to $F\neq 0$\ for $t>0$. After the
quench, the initial state $\left\vert \Psi \left( 0\right) \right\rangle $\
evolves according to the new Hamiltonian $\left\vert \Psi \left( t\right)
\right\rangle =e^{-iHt}\left\vert \Psi \left( 0\right) \right\rangle $. In
general, in presence of the external field, one would think that there can
exist a threshold beyond which the initial correlated particles state $%
\left\vert \Psi \left( 0\right) \right\rangle $\ should be pumped into the
scattering band and lose their pair correlation. The corresponding transfer
rate from the initial correlated particles state to final uncorrelated
states gets larger with increase of the field strength. However, after a
field quench, the dynamics of the correlated particles state will exhibit a
periodic behavior characterized by two distinct behaviors, owing to the
dramatic change of the spectrum from the multi avoided crossings to unmixed
region in a small range of the external field. This point will be presented
in the following.

Before starting the investigation of the dynamics of the two correlated
particles state, we would like to study the distribution of $\left\vert \Psi
\left( 0\right) \right\rangle $ in the energy space for the quenched systems
with and without multi avoided crossing regions. The corresponding spectrums
of the concerned two systems are plotted in Fig. \ref{fig4}. For the
quenched system with multi avoided crossings, the distribution of initial
correlated state $\left\vert \Psi \left( 0\right) \right\rangle $ mainly
concentrates on the mixed region and therefore refers to many energy levels.
This is due to the fact that the correlation exchange takes place among the
involved correlated state and uncorrelated band through the avoided
crossing. On the contrary, when the sudden change of the field strength $F$
is far away from the avoided crossing region, the wave functions of two
types of states are not mixed, which suppress the tunneling from the
correlated to uncorrelated states. This leads to the fact that the
corresponding eigenstates remain their correlation. Therefore, in contrast
to the case of the quenched system with multi avoided crossing regions, the
initial state $\left\vert \Psi \left( 0\right) \right\rangle $ is mainly
distributed on the correlated states and contains less energy levels, which
are equally spaced with an approximately constant spacing given by $F$. As a
comparison, we show in Fig. \ref{fig6} the main distribution of the initial
correlated wave packet $\left\vert \Psi \left( 0\right) \right\rangle $ in
the energy space of $H$ for two typical cases. It can be observed that the
energy distribution related to the quenched system with multi avoided
crossing regions involves more energy levels comparing with that of the
system with the unmixed region, which is in agreement with our analysis.
This transition of the distribution can not only reflect the change of the
energy spectrum but also result in the entirely different dynamical
behaviors of two such quenched systems. Thus it can be presumable that the
dynamics of the evolved state $\left\vert \Psi \left( t\right) \right\rangle
$ refered to the quenched system without multi avoided crossings will
exhibit BO with period $t_{B}=2\pi /F$ and hold its correlation, which is
analogous to a single particle motion subjected to an external field. On the
other hand, for the evolved state $\left\vert \Psi \left( t\right)
\right\rangle $ govern by the quenched system with multi avoided crossings,
the initial correlated particles will be disassociated and lose the
correlation between the two particles through the avoided crossing.

In order to characterize both the above two typical behaviors, as in Sec. %
\ref{sec_Machanism}, we introduce transfer rate $\overline{T}\left( t\right)
$ to measure how much probability of the evolved state will remain in the
bound states%
\begin{equation}
\overline{T}\left( t\right) =\sum_{\sigma ,K}\left\vert \left\langle \psi
_{K}^{\sigma }\right\vert e^{-iHt}\left\vert \Psi \left( 0\right)
\right\rangle \right\vert ^{2}.
\end{equation}%
We plot Fig. \ref{fig7} to verify and demonstrate the above analysis. For
the evolved state mainly placed on the multi avoided crossing regions, we
find that $\overline{T}\left( t\right) $ will tend to be $0$ accompanied by
the irreversible increase of $\overline{r}\left( t\right) $ as time goes on.
This indicates that the initial two correlated particles lose their
correlation, which can be seen from the Fig. \ref{fig7}(b). On the contrary,
in Fig. \ref{fig7}(a), the correlation of the two particles remains strong
and the corresponding $\overline{T}\left( t\right) $ tends to be a steady
value $0.93$ in a long time scale, which accords with our analytical
predictions. Furthermore, the step decrease of $\overline{T}\left( t\right) $
as a function of time in Fig. \ref{fig7}(b) can be explained from another
perspective as follows: After a field quench, the presence of the external
field can not only drive the correlated wave packet to move along the bound
band but also give rise to the Zener tunneling between the bound and
scattering band of $H_{0}$. The Zener tunnelling takes place almost
exclusively when the momentum of the wave packet reaches to the point of
center momentum $K=0$, which is either at the top of the bound band or at
the bottom of the scattering band. In this case, the correlated pair can be
partially pumped into the upper scattering band accompanied by the
destruction of correlation and the increase of energy. At this time, the
evolved state is a mixed state composed of correlated and scattering states.
Over a period of time, the scattering and correlated components of the
evolved state once again arrive at the transition point simultaneously with
a characteristic period given by $t_{B}=2\pi /F$. Then the two components of
the evolved state exchange the energy and correlation from each other due to
the Zener tunneling. Through a number of these processes, the $\overline{T}%
\left( t\right) $ tends to be a steady value. The corresponding time $t_{f}$%
\ can be termed as quench relaxation time, which deeply depends on the
distribution of the initial state in the energy space. To further understand
the field-induced energy variation arising from the Zener tunneling from the
bound band to scattering band, we study the average energy of $H_{0}$, which
can be defined by
\begin{equation}
\overline{E}\left( t\right) =\left\langle \Psi \left( t\right) \right\vert
H_{0}\left\vert \Psi \left( t\right) \right\rangle .
\end{equation}%
As comparison, the average energy $\overline{E}\left( t\right) $ as a
function of time for two typical quenched systems is plotted Fig. \ref{fig7}%
(a) and (b). In Fig. \ref{fig7}(a), the average energy of the evolved state
oscillates around the strength of on-site interaction $U=-6.24$ preserving
the constancy of the average interval energy. The corresponding oscillation
period is given by $t_{B}=2\pi /F$. This implies that the two correlated
particles undergo a BO, which is in accordance with our previous analysis.
While for the dynamics of correlated particles in the quenched system with
multi avoided crossing regions, the average energy of the evolved state is
first around the strength of on-site interaction $U=-6.24$ and then
oscillates around $0$, which indicates\ that the evolved state is completely
pumped into the scattering region with the increase of the average energy of
two particles. This can be seen from Fig. \ref{fig7}(b). On the other hand,
when the quench field is switched off at time $t_{f}$, the time evolution of
the state $\left\vert \Psi \left( t_{f}\right) \right\rangle $\ is driven by
the Hamiltonian $H_{0}$\ leading to a constant average energy. At this time,
the state $\left\vert \Psi \left( t_{f}\right) \right\rangle $\ are mainly
distributed on the scattering band of $H_{0}$\ owing to the increase of the
average distance $\overline{r}$, which indicates that the quench process
accomplishes the energy transfer from the field to particles. This is
analogous to the electrothermal effect in the context of classical physics,
which describes the interconversion of thermal energy difference and
electrical potential. Therefore, such a quench process, in which the energy
of the two particles is elevated by the external field, can be termed as a
quantum electrothermal effect. Furthermore, one can extend this result to
the system consisting of multi correlated pairs in which the pair-pair
interaction is neglected. The\ quench field can elevate the total energy of
the particles associated with the disassociation of the pairs in the
coordinate space. This process is reflected by the characteristic
temperature of the system, which can be measured in experiments.

Now we turn to study the sensitivity of the transition between the
correlated and uncorrelated particles states after a field quench. In the
previous section, we have pointed out that a slight variation of the
external field can trigger the periodic occurrence of the multi avoided
crossing regions, which can affect the stability of correlated pair. Thus
the dynamics of correlated pair is sensitive to the external field and can
characterize the change of the energy spectrum. Based on this analysis, we
investigate the time evolution of the initial state $\left\vert \Psi \left(
0\right) \right\rangle $\ by suddenly quenching the field strength from an
initial field-free value to a different value $F(t>0)$. And we employ long
time-scale transfer rate $\overline{T}_{t_{f}}\left( F\right) $\ to measure
the sensitivity of the correlated-pair dynamics to a\ quench field, which
can be described by%
\begin{equation}
\overline{T}_{t_{f}}\left( F\right) =\sum_{\sigma ,K}\left\vert \left\langle
\psi _{K}^{\sigma }\right\vert e^{-iH\left( F\right) t_{f}}\left\vert \Psi
\left( 0\right) \right\rangle \right\vert ^{2},
\end{equation}%
where $t_{f}$\ is the quench relaxation time. At this time, the $\overline{T}%
_{t_{f}}\left( F\right) $\ tends to be a steady value. We plot the quantity $%
\overline{T}_{t_{f}}\left( F\right) $\ as a function of quench field $F$\ in
Fig. \ref{fig8}. It can be observed that the long time-scale transfer rate $%
\overline{T}_{t_{f}}\left( F\right) $\ from an initial correlated state to
final uncorrelated particle states is sensitive to the quench field strength
and exhibits a periodic behavior. And the maximum of $\overline{T}%
_{t_{f}}\left( F\right) $\ is corresponding to the initial correlated pair
placed on the unmixed region of the\ quenched system. While the minimum of $%
\overline{T}_{t_{f}}\left( F\right) $\ denotes the correlated-pair dynamics
in the quenched system with multi avoided crossings. The numerical results
clearly show that the $\overline{T}_{t_{f}}\left( F\right) $\ is a good
indicator of the transition between systems with and without multi avoided
crossing regions induced by the external field. This may also provide a
method to identify a subtle change of the system parameter in experiments.

\section{Summary}

\label{sec_Summary} In this work, we have theoretically investigated the
coherent motion of two correlated particles driven by an external field in
the framework of the one-dimensional Hubbard model. We have shown that the
particle-particle interaction can induce BP states, exhibiting short-range
quantum correlations. The spectrum comprises two Bloch bands,
uncorrelated-particle and BP bands. The presence of the external field can
not only changes the energy spectrum from the continuous to discrete but
also bring about the mixture among the correlated state and multi
uncorrelated states, which is accompanied with the exchange of the energy
and correlation between the involved two types of states. We have also shown
that a small change of the external field can induce periodic emergence of
the multi avoided crossing regions leading to the transition from the
initial correlated state to final uncorrelated states. The corresponding
transfer rate is sensitive to the quench field strength and exhibits a
periodic behavior. This process is irreversible and accomplishes the energy
transfer from the field to particles, which could be the mechanism of the
quantum electrothermal effect. It is envisaged that the sensitivity of the
correlated-pair dynamics to an external field can be applied to detect a
subtle variation of system parameter, stimulating further theoretical and
experimental investigations on the dynamics of strongly interacting
particles in a variety of driven cold-atom systems.

\acknowledgments We acknowledge the support of the National Basic Research
Program (973 Program) of China under Grant No. 2012CB921900 and the CNSF
(Grant No. 11374163). X. Z. Zhang is supported by PhD research startup
foundation of Tianjin Normal University under Grant No. 52XB1415.

\end{document}